\documentclass[12pt,a4paper]{article}

\usepackage{amsmath}
\usepackage{amssymb}
\usepackage{amsfonts}
\usepackage{amsthm}
\usepackage{mathtools}
\usepackage{bm}

\usepackage{physics}
\usepackage{slashed}

\usepackage{graphicx}
\usepackage{float}
\usepackage{subfig}
\usepackage{caption}
\usepackage{booktabs}
\usepackage{multirow}

\usepackage[a4paper,margin=1in]{geometry}
\usepackage{setspace}
\usepackage{indentfirst}

\usepackage[colorlinks=true,
            linkcolor=blue,
            citecolor=blue,
            urlcolor=blue]{hyperref}

\usepackage[numbers,sort&compress]{natbib}

\usepackage{array}
\usepackage{longtable}

\usepackage{textcomp}
\usepackage{gensymb}

\usepackage{xcolor}
\usepackage{enumitem}

\usepackage{subcaption}

\title{
\textbf{
When Does Leptogenesis Survive Lepton Flavor Violation Constraints?\\
High- and Low-Scale Realizations in the Scotogenic Model
}
}

\author{
Avinanda Chaudhuri$^{1}$\thanks{aviphys@gmail.com}
\\[0.5cm]
\small $^{1}$Department of Physics, \\ Brahmananda Keshab Chandra College\\, 111/2, B. T. Road, Kolkata - 700108, India
}

\date{20.04.2026}

\begin{document}

\maketitle

\begin{abstract}

We investigate the interplay between lepton flavor violation (LFV) and leptogenesis in the minimal scotogenic model, comparing high-scale hierarchical leptogenesis and low-scale resonant leptogenesis within a unified Casas--Ibarra framework. Since the same Yukawa couplings simultaneously govern radiative neutrino mass generation, charged LFV processes, and the CP asymmetry required for baryogenesis, strong phenomenological correlations arise. We show that high-scale leptogenesis remains naturally viable due to the effective decoupling between LFV and baryogenesis, while low-scale resonant leptogenesis is strongly constrained by the MEG bound on $\mu \rightarrow e\gamma$. Nevertheless, we identify a narrow but nonvanishing resonant window where successful baryogenesis, controlled washout, and LFV safety coexist simultaneously. In particular, we obtain fully allowed benchmark points characterized by quasi-degenerate heavy fermions, resonantly enhanced CP asymmetry, and suppressed flavor violation through Casas--Ibarra phase alignment. 

\end{abstract}

\tableofcontents

\newpage


\section{Introduction}

The baryon asymmetry of the Universe (BAU)~\cite{PDG2024, Planck2018} and the tiny but non-zero masses of neutrinos, as confirmed by various experiments, are among the most compelling evidence for the existence of new physics beyond the Standard Model (SM) of particle physics. Among the viable mechanisms to explain the observed BAU, leptogenesis is one of the most promising.

The simplicity and close connection of leptogenesis~\cite{FukugitaYanagida1986} with low-energy neutrino physics make it an experimentally testable scenario in future collider and neutrino experiments.

In its most standard formulation, leptogenesis is closely related to the Type-I seesaw mechanism~\cite{Minkowski1977, Yanagida1979, GellMann1979, Mohapatra1980, Schechter1980}, which employs two or more sterile right-handed neutrinos (RHNs), denoted by $N_i$, with large Majorana masses $M_i$ to explain the tiny masses of neutrinos of SM.

These heavy RHNs, produced through scattering processes in the early Universe, generate a net lepton asymmetry through their CP-violating out-of-equilibrium decays. This generated lepton asymmetry is then partially converted into the observed baryon asymmetry through sphaleron processes. For review articles on leptogenesis, see~\cite{Dev2018review, Drewes2018, Dev2018resonant, Biondini2018, Chun2018, Hagedorn2018}.

An intrinsic limitation of normal thermal leptogenesis is that the involved RHN mass scales are very high. An absolute lower bound on the mass of the lightest RHN is approximately

\[
M_{1} \gtrsim 10^{9}\,\text{GeV}
\]

for successful baryogenesis, which was first pointed out by Davidson and Ibarra~\cite{DavidsonIbarra}.

Such high mass scales are very challenging to directly probe the dynamics of leptogenesis in future collider experiments. Motivated by these concerns, several alternative possibilities were presented that managed to generate the observed BAU at a much lower scale (TeV) of RHN masses~\cite{Asaka2005, Pilaftsis2004, Hambye2016, Drewes2018ARS, Klaric2021, Jukkala2021, Racker2023, Canetti2013, Ghiglieri2019, Ghiglieri2020, Tao1996, Ma2006PRD, Ma2006MPLA}.

This situation changes substantially if, together with the RHNs, a second inert scalar doublet is added to the model and the new fields are required to be odd under a discrete $Z_2$ symmetry, while all SM particles remain even under this discrete symmetry construction.

This model, known as the Scotogenic model~\cite{ Ma2006PRD}, is arguably the simplest and most studied framework to explain simultaneously the smallness of neutrino masses, leptogenesis, and dark matter. In this framework, neutrino masses are generated radiatively at the one-loop level through the exchange of an inert scalar doublet and heavy Majorana fermions.  Remarkably, the same Yukawa couplings responsible for neutrino mass generation also govern charged lepton flavor violating processes and the CP asymmetry required for leptogenesis, leading to a highly predictive and tightly constrained framework.

This intimate connection  makes the scotogenic model particularly suitable for exploring the correlation between flavor physics and cosmology. In particular, the radiative origin of neutrino masses implies that the branching ratio of processes such as
\begin{equation}
\mu \rightarrow e\gamma
\end{equation}
depends directly on the same Yukawa structure entering the neutrino mass matrix and leptogenesis dynamics. Consequently, the stringent upper bound from the MEG experiment~\cite{MEG2016,MEGII2025} places severe restrictions on the parameter space relevant for successful baryogenesis.

A central question therefore emerges:

\textit{Can leptogenesis survive the strong constraints imposed by lepton flavor violation in the scotogenic model?}

This question becomes particularly nontrivial when one compares two distinct realizations of leptogenesis:

\begin{enumerate}
    \item \textbf{High-scale hierarchical leptogenesis}, where the heavy Majorana fermions are strongly hierarchical and the CP asymmetry arises from standard decay asymmetries;

    \item \textbf{Low-scale resonant leptogenesis}, where quasi-degenerate heavy fermions lead to resonant enhancement of the CP asymmetry, allowing successful baryogenesis even at comparatively low scales.
\end{enumerate}

Although both possibilities exist in the literature, a systematic comparison between high-scale and low-scale realizations within the same Casas--Ibarra reconstruction framework remains comparatively limited. In particular, the viability of low-scale resonant leptogenesis under current LFV bounds requires a dedicated numerical reassessment.

In this work, we performed a comprehensive study of the possible connection between LFV and leptogenesis in the minimal scotogenic model using the full Casas--Ibarra parameterization of the Yukawa sector. We include both hierarchical and resonant leptogenesis scenarios and solve the relevant Boltzmann equations to determine the final baryon asymmetry. Special emphasis is placed on the competition between resonant CP enhancement and LFV suppression. The LFV–leptogenesis correlation studied here is largely independent
of the precise dark matter realization.

Our analysis reveals a clear scale-dependent structure. We find that high-scale leptogenesis remains naturally viable, since LFV and baryogenesis can effectively decouple through flavor alignment and Casas--Ibarra phase cancellations. In contrast, low-scale resonant leptogenesis is strongly constrained because the same Yukawa enhancement required for resonant CP asymmetry tends to violate the bound $\mu \to e\gamma$ and induces excessive washout. Nevertheless, contrary to the common expectation of complete exclusion, we identify a narrow but non vanishing resonant window in which successful baryogenesis, controlled washout, and LFV safety coexist simultaneously. This surviving region is characterized by quasi-degenerate heavy fermions, finely tuned Casas--Ibarra phases, and suppressed flavor-violating amplitudes.

The main result of this work may be summarized as

\textit{High-scale leptogenesis survives naturally, whereas low-scale resonant leptogenesis survives only in a highly restricted LFV-safe resonant strip.}

This provides a strong phenomenological target for future LFV searches and offers a predictive framework linking neutrino physics and baryogenesis.

The paper is organized as follows. In section 2, we present the minimal scotogenic framework and derive the one-loop neutrino mass matrix. Section 3 discusses the Casas--Ibarra reconstruction of the Yukawa couplings and the role of complex orthogonal rotations. In section 4, we present the formalism for LFV observables and leptogenesis, including CP asymmetry, washout dynamics, and Boltzmann evolution. Section 5 describes the numerical setup and scanning strategy. Finally, section 6 summarizes our conclusions.


\section{The Minimal Scotogenic Framework}

The minimal scotogenic model extends the Standard Model (SM) by introducing three singlet Majorana fermions $N_i$ $(i=1,2,3)$ and one additional scalar doublet $\eta$, all of which are odd under an exact discrete $Z_2$ symmetry, while all Standard Model fields remain $Z_2$-even. This discrete symmetry simultaneously forbids the tree-level Dirac neutrino mass term and guaranties the stability of the lightest $Z_2$-odd particle.

The particle content relevant for neutrino mass generation is summarized as follows:
\begin{align}
L_\alpha &= 
\begin{pmatrix}
\nu_\alpha \\
\ell_\alpha
\end{pmatrix}_L
\sim (2,-\tfrac{1}{2},+),
\\[2mm]
H &= 
\begin{pmatrix}
H^+ \\
H^0
\end{pmatrix}
\sim (2,\tfrac{1}{2},+),
\\[2mm]
\eta &= 
\begin{pmatrix}
\eta^+ \\
\eta^0
\end{pmatrix}
\sim (2,\tfrac{1}{2},-),
\\[2mm]
N_i &\sim (1,0,-),
\end{align}
where the last entry denotes the $Z_2$ parity.

The neutral component of the inert scalar doublet is decomposed as
\begin{equation}
\eta^0
=
\frac{1}{\sqrt{2}}
\left(
\eta_R + i \eta_I
\right),
\end{equation}
where $\eta_R$ and $\eta_I$ denote the CP-even and CP-odd neutral scalars, respectively.

Importantly, unlike the SM Higgs doublet, the inert doublet does not acquire a vacuum expectation value (VEV),
\begin{equation}
\langle \eta \rangle = 0,
\end{equation}
which preserves the exact $Z_2$ symmetry after electroweak symmetry breaking.

\subsection{Scalar Potential}

The most general renormalizable scalar potential~\cite{Deshpande1978} consistent with the gauge symmetry and the exact $Z_2$ symmetry is given by
\begin{align}
V(H,\eta)
&=
\mu_1^2 |H|^2
+
\mu_2^2 |\eta|^2
+
\lambda_1 |H|^4
+
\lambda_2 |\eta|^4
\nonumber\\
&\quad
+
\lambda_3 |H|^2 |\eta|^2
+
\lambda_4 |H^\dagger \eta|^2
+
\frac{\lambda_5}{2}
\left[
(H^\dagger \eta)^2 + \text{h.c.}
\right].
\end{align}

After electroweak symmetry breaking, the SM Higgs doublet develops the vacuum expectation value
\begin{equation}
\langle H \rangle
=
\frac{1}{\sqrt{2}}
\begin{pmatrix}
0 \\
v
\end{pmatrix},
\qquad
v \simeq 246~\text{GeV},
\end{equation}
The physical masses of the inert scalar components are then given by
\begin{align}
m_{\eta^\pm}^2
&=
\mu_2^2
+
\frac{\lambda_3 v^2}{2},
\\[2mm]
m_{\eta_R}^2
&=
\mu_2^2
+
\frac{v^2}{2}
\left(
\lambda_3 + \lambda_4 + \lambda_5
\right),
\\[2mm]
m_{\eta_I}^2
&=
\mu_2^2
+
\frac{v^2}{2}
\left(
\lambda_3 + \lambda_4 - \lambda_5
\right).
\end{align}

The parameter $\lambda_5$ plays a particularly important role since it controls the mass splitting between the neutral inert scalars,
\begin{equation}
m_{\eta_R}^2
-
m_{\eta_I}^2
=
\lambda_5 v^2.
\end{equation}

This splitting is essential for generating nonzero neutrino masses at one loop. In the limit
$\lambda_5 \to 0,$ lepton number is restored and neutrino masses vanish.

\subsection{Yukawa Sector and Majorana Masses}

The leptonic Yukawa interactions relevant for neutrino mass generation are given by
\begin{equation}
\mathcal{L}_Y
=
Y_{\alpha i}
\,
\overline{L_\alpha}
\,
\widetilde{\eta}
\,
N_i
+
\text{h.c.},
\end{equation}
where $\widetilde{\eta} = i \sigma_2 \eta^*,$ and $Y_{\alpha i}$ denotes the neutrino Yukawa coupling matrix.

The heavy singlet fermions possess Majorana mass terms
\begin{equation}
\mathcal{L}_M
=
\frac{1}{2}
M_i
\,
\overline{N_i^c}
N_i
+
\text{h.c.},
\end{equation}
which explicitly violate lepton number and play a crucial role in both neutrino mass generation and leptogenesis.

Because of the exact $Z_2$ symmetry, the standard Dirac neutrino mass term
$\overline{L}\widetilde{H}N $
is forbidden.

\subsection{One-Loop Neutrino Mass Matrix}

The active neutrino masses are generated at the one-loop level through the exchange of the inert neutral scalars $(\eta_R,\eta_I)$ and the heavy Majorana fermions $N_i$.

The resulting neutrino mass matrix~\cite{Kubo2006,Restrepo2013, Merle2015,Ahriche2018} is given by
\begin{equation}
(M_\nu)_{\alpha\beta}
=
\sum_{k=1}^{3}
\frac{Y_{\alpha k} Y_{\beta k} M_k}{32\pi^2}
\left[
\frac{m_{\eta_R}^2}
{m_{\eta_R}^2 - M_k^2}
\ln
\left(
\frac{m_{\eta_R}^2}{M_k^2}
\right)
-
\frac{m_{\eta_I}^2}
{m_{\eta_I}^2 - M_k^2}
\ln
\left(
\frac{m_{\eta_I}^2}{M_k^2}
\right)
\right].
\label{eq:full_neutrino_mass}
\end{equation}

It is convenient to define the loop function
\begin{equation}
\Lambda_k
=
\frac{M_k}{32\pi^2}
\left[
\frac{m_{\eta_R}^2}
{m_{\eta_R}^2 - M_k^2}
\ln
\left(
\frac{m_{\eta_R}^2}{M_k^2}
\right)
-
\frac{m_{\eta_I}^2}
{m_{\eta_I}^2 - M_k^2}
\ln
\left(
\frac{m_{\eta_I}^2}{M_k^2}
\right)
\right],
\end{equation}

such that the neutrino mass matrix takes the compact form. Here \(m_R\) and \(m_I\) are the CP-even and CP-odd neutral inert scalar masses, respectively, and \(\lambda_5\) controls their mass splitting.
\begin{equation}
M_\nu
=
Y
\,
\Lambda
\,
Y^T,
\qquad
\Lambda
=
\text{diag}
\left(
\Lambda_1,
\Lambda_2,
\Lambda_3
\right).
\label{eq:compact_neutrino_mass}
\end{equation}

Equation~(\ref{eq:compact_neutrino_mass}) is the central relation of the scotogenic model. It directly connects neutrino masses to the same Yukawa couplings responsible for charged lepton flavor violation and leptogenesis. This unified structure makes the model highly predictive.


\section{Casas--Ibarra Reconstruction of the Yukawa Sector}

Since the low-energy neutrino mass matrix is experimentally constrained by oscillation data~\cite{SuperK:1998, SNO:2002, KamLAND:2003, DayaBay:2012, RENO:2012, DoubleChooz:2012,T2K:2011, NOvA:2016, KATRIN:2024}, it is convenient to reconstruct the Yukawa couplings using the Casas--Ibarra parametrization~\cite{Casas:2001sr}, which allows a systematic exploration of the viable parameter space while preserving consistency with neutrino data.

\subsection{Neutrino inputs and complex orthogonal matrix}

The light neutrino mass matrix is diagonalized by the Pontecorvo--Maki-Nakagawa-Sakata (PMNS) matrix~\cite{PMNS:Pontecorvo, PMNS:MNS, PMNS:Review},
\begin{equation}
U_{\rm PMNS}^T
\,
M_\nu
\,
U_{\rm PMNS}
=
D_\nu,
\end{equation}
where
$
D_\nu = \text{diag}\left(m_1,\,m_2,\,m_3\right)$ contains the physical light neutrino masses.

The PMNS matrix is parametrized as
\begin{equation}
U_{\rm PMNS}
=
\begin{pmatrix}
1 & 0 & 0
\\
0 & c_{23} & s_{23}
\\
0 & -s_{23} & c_{23}
\end{pmatrix}
\begin{pmatrix}
c_{13} & 0 & s_{13}e^{-i\delta}
\\
0 & 1 & 0
\\
-s_{13}e^{i\delta} & 0 & c_{13}
\end{pmatrix}
\begin{pmatrix}
c_{12} & s_{12} & 0
\\
-s_{12} & c_{12} & 0
\\
0 & 0 & 1
\end{pmatrix}
\begin{pmatrix}
1 & 0 & 0
\\
0 & e^{i\alpha/2} & 0
\\
0 & 0 & e^{i\beta/2}
\end{pmatrix},
\end{equation}
where
$
c_{ij} = \cos\theta_{ij},
s_{ij} = \sin\theta_{ij},
$
and $\delta$, $\alpha$, and $\beta$ denote the Dirac and Majorana CP phases.

In this work we have used the latest global-fit neutrino oscillation parameters~\cite{NuFIT:2024} assuming normal hierarchy (NH) without any loss of generality.

Starting from the compact scotogenic neutrino mass relation
\begin{equation}
M_\nu
=
Y
\,
\Lambda
\,
Y^T,
\end{equation}
the Yukawa matrix can be reconstructed as
\begin{equation}
Y
=
U_{\rm PMNS}
\,
\sqrt{D_\nu}
\,
R
\,
\sqrt{\Lambda^{-1}},
\label{eq:casas_ibarra}
\end{equation}
where $R$ is a complex orthogonal matrix satisfying
\begin{equation}
R^T R = \mathbb{I}.
\end{equation}

Equation~(\ref{eq:casas_ibarra}) provides the exact generalization of the Casas--Ibarra parametrization to the scotogenic framework and allows one to reconstruct Yukawa couplings consistent with neutrino oscillation data.

\subsection{Role of the complex angles}

The matrix $R$ contains the unconstrained high-energy degrees of freedom that are not fixed by low-energy neutrino data. It is parametrized in terms of three generally complex angles,
\begin{equation}
\theta_i
=
x_i + i y_i,
\qquad
(i=1,2,3),
\end{equation}
where $x_i$ and $y_i$ are real parameters.

A convenient explicit form is
\begin{equation}
R
=
R_{23}
R_{13}
R_{12},
\end{equation}

with
\begin{equation}
R_{12}
=
\begin{pmatrix}
\cos\theta_1 & -\sin\theta_1 & 0
\\
\sin\theta_1 & \cos\theta_1 & 0
\\
0 & 0 & 1
\end{pmatrix},
\end{equation}
\begin{equation}
R_{13}
=
\begin{pmatrix}
\cos\theta_2 & 0 & -\sin\theta_2
\\
0 & 1 & 0
\\
\sin\theta_2 & 0 & \cos\theta_2
\end{pmatrix},
\end{equation}
and
\begin{equation}
R_{23}
=
\begin{pmatrix}
1 & 0 & 0
\\
0 & \cos\theta_3 & -\sin\theta_3
\\
0 & \sin\theta_3 & \cos\theta_3
\end{pmatrix}.
\end{equation}

Because the angles are complex,
\begin{equation}
\sin(x+iy)
=
\sin x \cosh y
+
i \cos x \sinh y,
\end{equation}
\begin{equation}
\cos(x+iy)
=
\cos x \cosh y
-
i \sin x \sinh y,
\end{equation}
This complex structure is crucial for the CP asymmetry parameter
\begin{equation}
\epsilon_i
\propto
\sum_{j\neq i}
\frac{
\mathrm{Im}
\left[
\left(
Y^\dagger Y
\right)^2_{ij}
\right]
}{
\left(
Y^\dagger Y
\right)_{ii}
},
\end{equation}
which vanishes for purely real Yukawa couplings. Therefore, the imaginary part of the Casas--Ibarra angle directly controls the source of leptogenesis and satisfies one of the Sakharov conditions~\cite{Sakharov:1967dj}, namely CP violation.

In the low-scale resonant regime, successful leptogenesis does not necessarily require unnaturally large Yukawa couplings. Instead, the enhancement originates from the quasi-degeneracy condition
$
M_2 \simeq M_1,
$
which resonantly amplifies the CP asymmetry. This allows one to explore neutrino masses,  baryogenesis, and LFV bounds without excessive fine-tuning of individual Yukawa entries.

In our analysis, we performed a full scan of complex angles $\theta_i$ and study how the competition between LFV suppression and enhancement of CP asymmetry determines the surviving parameter space for both high-scale and low-scale leptogenesis.

This unified role makes the numerical exploration of the complex $R$-matrix parameter space essential for identifying viable benchmark regions. 
%


\section{Lepton Flavor Violation and Leptogenesis Formalism}


In this section, we present the formalism relevant for LFV observables, CP asymmetry generation, washout dynamics, and the Boltzmann evolution leading to the final baryon asymmetry of the Universe.

\subsection{Charged Lepton Flavor Violation}

Among all LFV observables, the radiative decay
\begin{equation}
\mu \rightarrow e\gamma
\end{equation}
provides the strongest and most constraining bound on the parameter space of the model. Since the branching ratio depends directly on the neutrino Yukawa couplings, the MEG bound strongly restricts the same parameters relevant for leptogenesis.

The branching ratio of the process is given by~\cite{ Toma:2013zsa}
\begin{equation}
\text{BR}(\mu \to e\gamma)
=
\frac{3\alpha_{\rm em}}
{64\pi G_F^2 m_{\eta^\pm}^4}
\left|
\sum_{k=1}^{3}
Y_{\mu k}
Y_{e k}^*
F_2
\left(
\frac{M_k^2}{m_{\eta^\pm}^2}
\right)
\right|^2,
\label{eq:muegamma}
\end{equation}
where $\alpha_{\rm em}$ is the electromagnetic fine-structure constant, $G_F$ is the Fermi constant, and the loop function is
\begin{equation}
F_2(x)
=
\frac{
1 - 6x + 3x^2 + 2x^3 - 6x^2 \ln x
}
{6(1-x)^4}.
\end{equation}
The current experimental upper bound from the MEG~\cite{MEGII2025} collaboration is
\begin{equation}
\text{BR}(\mu \to e\gamma)
<
4.2 \times 10^{-13},
\label{eq:meg_bound}
\end{equation}
which constitutes the dominant LFV constraint in our numerical analysis.

Although other LFV channels such as
$
\mu \to 3e,
\mu\text{-}e \text{ conversion},
\tau \to \ell\gamma
$
also provide useful complementary information, they are generally subdominant compared to the strong sensitivity of $\mu \to e\gamma$. We therefore focus primarily on the MEG bound.
\subsection{Hierarchical  and resonant Leptogenesis}

In the high-scale regime where the heavy Majorana fermions satisfy
\begin{equation}
M_1 \ll M_2 \ll M_3,
\end{equation}
the baryon asymmetry is generated through the out-of-equilibrium decay of the lightest singlet fermion $N_1$.

The CP asymmetry parameter is defined as
\begin{equation}
\epsilon_1
=
\frac{
\Gamma(N_1 \to \ell \eta)
-
\Gamma(N_1 \to \bar{\ell}\,\eta^\dagger)
}
{
\Gamma(N_1 \to \ell \eta)
+
\Gamma(N_1 \to \bar{\ell}\,\eta^\dagger)
}.
\end{equation}

For hierarchical masses, the leading contribution is
\begin{equation}
\epsilon_1
=
\frac{1}
{8\pi (Y^\dagger Y)_{11}}
\sum_{j\neq 1}
\text{Im}
\left[
\left(
(Y^\dagger Y)_{1j}
\right)^2
\right]
f
\left(
\frac{M_j^2}{M_1^2}
\right),
\label{eq:epsilon_hierarchical}
\end{equation}
where the loop function $f(x)$ contains both vertex and self-energy corrections.
In this regime, successful leptogenesis typically requires DI bound
$
M_1
\gtrsim
10^9~\text{GeV},
$
unless special flavor effects or resonant enhancement are present.
In the low-scale regime, successful baryogenesis may still occur if two heavy Majorana fermions are quasi-degenerate,
\begin{equation}
M_2
\simeq
M_1.
\end{equation}

The degree of degeneracy is characterized by
\begin{equation}
\Delta
=
\frac{M_2 - M_1}{M_1}.
\label{eq:delta}
\end{equation}

In the quasi-degenerate regime,
the self-energy contribution dominates the CP asymmetry and must be treated
with the proper resonant regulator. We employ the Pilaftsis–Underwood~\cite{Pilaftsis2004} regulator to consistently treat the quasi-degenerate limit and avoid the spurious divergence of the naive self-energy expression. The CP asymmetry generated from the decay
of the lightest singlet fermion \(N_1\) can then be written as
\begin{equation}
\epsilon_1
=
\frac{
\mathrm{Im}
\left[
\left(
Y^\dagger Y
\right)_{12}^{2}
\right]
}{
\left(
Y^\dagger Y
\right)_{11}
\left(
Y^\dagger Y
\right)_{22}
}
\,
\frac{
\left(
M_2^2-M_1^2
\right)
M_1
\Gamma_2
}{
\left(
M_2^2-M_1^2
\right)^2
+
M_1^2
\Gamma_2^2
},
\label{eq:resonant_cp}
\end{equation}
where
\begin{equation}
\Gamma_2
=
\frac{
M_2
}{
8\pi
}
\left(
Y^\dagger Y
\right)_{22}
\left(
1-\eta_2
\right)^2
\end{equation}
is the total decay width of \(N_2\), with
\begin{equation}
\eta_2
=
\frac{
m_\eta^2
}{
M_2^2
}.
\end{equation}

The regulator term
\[
\left(
M_2^2-M_1^2
\right)^2
+
M_1^2\Gamma_2^2
\]
prevents the unphysical divergence in the exact degenerate limit and shows
explicitly that maximal enhancement occurs when the mass splitting becomes
comparable to the decay width,
\[
M_2-M_1
\sim
\Gamma_2.
\]
This provides the physical origin of successful low-scale resonant leptogenesis
in the minimal scotogenic model.
\subsection{The Boltzmann equation and washout parameter}

The evolution of the heavy fermion abundance and the generated lepton asymmetry is described by the Boltzmann equations. Since the low-scale resonant regime considered in this work may lie in the
two- or three-flavor leptogenesis domain, a fully flavored Boltzmann treatment
would in principle be required for complete precision. However, for the purpose
of identifying the dominant parametric dependence and the interplay between LFV,
CP asymmetry, and washout effects, we work in the one-flavor approximation.
This captures the main structure of the viable parameter space, while a fully
flavored analysis is expected to modify the efficiency factor quantitatively
without changing the qualitative conclusions of the present study.

Defining
\begin{equation}
z
=
\frac{M_1}{T},
\end{equation}
the relevant coupled equations are~\cite{Buchmuller:2004nz, Davidson:2008bu, Hambye:2016sby}
\begin{align}
\frac{dY_{N_1}}{dz}
&=
-
D(z)
\left(
Y_{N_1}
-
Y_{N_1}^{\rm eq}
\right),
\\[2mm]
\frac{dY_{B-L}}{dz}
&=
-
\epsilon_1
D(z)
\left(
Y_{N_1}
-
Y_{N_1}^{\rm eq}
\right)
-
W(z)
Y_{B-L},
\label{eq:boltzmann}
\end{align}
Here, \(D(z)\) denotes the decay term, \(W(z)\) represents the washout term, \(Y_{N_1}^{\rm eq}\) is the equilibrium abundance of the lightest right-handed neutrino \(N_1\), and \(Y_{B-L}\) corresponds to the generated lepton asymmetry.

The equilibrium abundance is given by
$
Y_{N_1}^{\rm eq}
=
\frac{
45
}{
4\pi^4 g_*
}
z^2
K_2(z),
$
where $K_2$ is the modified Bessel function and
$
g_*
=
106.75
$
is the effective number of relativistic degrees of freedom.
The decay parameter
\(
K_1
\)
determines the overall strength of washout, while the washout function
\(
W(z)
\)
describes its temperature evolution in the Boltzmann equations, with
\(
z = M_1/T
\).
In the strong washout regime,
\(
W(z)
\)
scales proportionally with
\(
K_1
\),
implying that larger Yukawa couplings enhance both the CP asymmetry and the washout simultaneously.

The decay parameter is defined as
\begin{equation}
K_1
=
\frac{\Gamma_{N_1}}{H(T=M_1)},
\end{equation}
where
\(
\Gamma_{N_1}
\)
is the decay width of the lightest right-handed neutrino and
\(
H(T=M_1)
\)
is the Hubble expansion rate evaluated at
\(
T=M_1
\).
For inverse decays, one approximately has
\begin{equation}
W(z)
\simeq
\frac{1}{4}
K_1
z^3
\mathcal{K}_1(z),
\end{equation}
where
\(
\mathcal{K}_1(z)
\)
is the modified Bessel function of the second kind. In addition to inverse decays, $\Delta L = 2$ scattering processes can also contribute
to the washout of the generated lepton asymmetry, particularly in the strong washout
regime. The total washout term can therefore be written as
\[
W^{\rm tot}
=
W
+
\Delta W,
\]
where $W $ denotes the inverse-decay contribution and $\Delta W$ represents
the washout from lepton-number-violating scatterings such as
\[
\ell \eta \leftrightarrow \bar{\ell}\eta^\dagger,
\qquad
\ell\ell \leftrightarrow \eta\eta.
\]

For moderate values of the Yukawa couplings and sufficiently large RHN masses,
the inverse-decay term remains dominant and $\Delta W$ provides only a subleading
correction. In the present analysis, we focus on this regime and use the one-flavor
Boltzmann approximation with $W_{\rm ID}$ as the leading washout source. Including
$\Delta L = 2$ effects would mainly modify the efficiency factor quantitatively,
without changing the qualitative conclusion regarding the tension between LFV
constraints and successful leptogenesis.
\subsection{Approximate upper bound on the CP asymmetry}

In the hierarchical regime where
\begin{equation}
M_1 \ll M_2 \ll M_3,
\end{equation}
the CP asymmetry generated from the decay of the lightest singlet fermion \(N_1\) is constrained by a Davidson--Ibarra type upper bound adapted to the scotogenic framework. Owing to the radiative origin of neutrino masses, the standard seesaw bound receives an additional enhancement factor proportional to \(1/\lambda_5\), reflecting the relation
\begin{equation}
M_\nu
\propto
\lambda_5 Y^2.
\end{equation}

An approximate upper limit can be written as
\begin{equation}
|\epsilon_1|
\lesssim
\frac{3\pi}{
4\lambda_5 v^2
}
\,
\xi_3
\,
(m_h-m_l)
\,
M_1,
\label{eq:eps_bound}
\end{equation}
where \(m_h\) and \(m_l\) denote the heaviest and lightest active neutrino masses, respectively, and \(\xi_3\) is the loop correction factor.

In the low-scale resonant regime with
$
M_2 \simeq M_1 \ll M_3.
\label{eq:resonantpair}
$
the self-energy contribution becomes resonantly enhanced and the effective CP asymmetry can significantly exceed the hierarchical bound of Eq.~(\ref{eq:eps_bound}). This allows successful baryogenesis even for comparatively small Yukawa couplings and TeV-scale singlet fermions, which constitutes the main phenomenological advantage of resonant leptogenesis in the minimal scotogenic model.

Although the complete neutrino mass reconstruction requires all three singlet fermions, the dominant contribution to low-scale leptogenesis typically arises from the quasi-degenerate pair
In this limit, the self-energy contribution involving \(N_2\) becomes resonantly enhanced, while the contribution from \(N_3\) is comparatively suppressed due to the larger mass hierarchy.
Therefore, the pair
\((N_1,N_2)\)
effectively controls the leptogenesis dynamics, while \(N_3\) is mainly involved in the reconstruction of neutrino masses.

Schematically, the dominant dependence may be expressed as
\begin{equation}
\epsilon_1
\propto
\frac{
(m_h-m_l)
}{
\lambda_5 v^2
}
\,
M_1
\,
\frac{
\sin(2x_{12})
\sinh(2y_{12})
}{
\delta
}
\,
\xi_2,
\label{eq:eps_analytic}
\end{equation}
This result shows that the imaginary part of the complex angle directly governs the generated baryon asymmetry.

For the benchmark points considered in our numerical analysis, we typically obtain
$
K_1 \gg 1,
$
which places the system safely in the strong washout regime. In this limit, the efficiency factor may be approximated as
\begin{equation}
\kappa_1(K_1)
\simeq
\frac{1}{
1.2\,K_1
\left[
\ln K_1
\right]^{0.8}
}.
\label{eq:kappa_strong}
\end{equation}

\vspace{0.3cm}
\subsection{Approximate analytical expression for the baryon asymmetry}
The final baryon-to-photon ratio is then given by
\begin{equation}
\eta_B
\simeq
-C\,
\epsilon_1\,
\kappa_1,
\end{equation}
where
\(C\sim10^{-2}\)
contains the sphaleron conversion and entropy dilution factor.

Substituting Eqs.~(\ref{eq:eps_analytic}) and
(\ref{eq:kappa_strong}),
we obtain the compact approximate expression
\begin{equation}
\eta_B
\simeq
C\,
\frac{
(m_h-m_l)
}{
\lambda_5 v^2
}
\,
M_1
\,
\frac{
\sin(2x_{12})
\sinh(2y_{12})
}{
\delta
}
\,
\frac{
1
}{
K_1
\left[
\ln K_1
\right]^{0.8}
}
\,
\xi_2.
\label{eq:etaB_master}
\end{equation}

This expression should be understood as a schematic scaling relation
rather than an exact analytical formula, summarizing the essential physics of low-scale resonant leptogenesis in the minimal scotogenic model.

A characteristic feature of the scotogenic model is the dependence on the scalar coupling
\(\lambda_5\).
Since the neutrino mass scale satisfies
\begin{equation}
M_\nu
\propto
\lambda_5\,Y^2,
\end{equation}
smaller values of
\(\lambda_5\)
require larger Yukawa couplings to reproduce oscillation data.

Naively, this enhances the CP asymmetry.
However, the same increase in Yukawa couplings also enlarges the washout parameter
\(K_1\),
thereby suppressing the efficiency factor.
As a result, decreasing
\(\lambda_5\)
does not always increase the final baryon asymmetry and may even reduce
\(\eta_B\)
in the strong washout regime.
This explains the behaviour observed in our
\(\epsilon_1\)--\(\lambda_5\)
and allowed-region plots.
\section{Numerical Analysis and Viable Parameter Space}

In this section, we perform a detailed numerical analysis of the minimal scotogenic model and identify the parameter regions simultaneously compatible with neutrino masses, successful leptogenesis, charged lepton flavor violation (LFV) bounds, and perturbativity. Our primary goal is to test the analytical expectations derived in the previous section and determine the viable benchmark regions in both the high-scale hierarchical regime and the low-scale quasi-degenerate regime.

The scan is performed over the parameters
\begin{align}
M_1,\quad \lambda_5,\quad |Y_{\alpha i}|,\quad 
\Delta \equiv \frac{M_2-M_1}{M_1},
\end{align}
together with the Casas--Ibarra complex angles controlling the CP violating structure of the Yukawa sector. 
Only points satisfying all theoretical and experimental constraints are classified as fully allowed.

\begin{table}[t]
\centering
\caption{Input parameter ranges used in the numerical scan. All points are required to satisfy neutrino oscillation data, perturbativity, vacuum stability, LFV bounds, and successful leptogenesis.}
\label{tab:scan_parameters}
\begin{tabular}{|c|c|c|}
\hline
Parameter & Scan Range & Description \\
\hline
$M_{1}$ & $10^{5} \text{--} 10^{13}\ \mathrm{GeV}$ 
& Lightest right-handed neutrino mass \\
\hline

$M_{2}$ & $M_{1}(1+\Delta)$ 
& Nearly degenerate second RH neutrino mass \\
\hline

$\Delta = \dfrac{M_{2}-M_{1}}{M_{1}}$ 
& $10^{-6} \text{--} 10^{-5}$ 
& Resonant mass splitting parameter \\
\hline

$\lambda_{5}$ 
& $10^{-6} \text{--} 10^{-2}$ 
& Inert doublet quartic coupling \\
\hline

$m_{\eta}$ 
& $500 \text{--} 3000\ \mathrm{GeV}$ 
& Inert scalar mass scale \\
\hline


$\theta_{1}$ 
& $0 \text{--} 2\pi$ 
& Casas--Ibarra angle \\
\hline

$\theta_{2}$ 
& $0 \text{--} 2\pi$ 
& Casas--Ibarra angle \\
\hline

$Y_{\alpha i}$ 
& perturbative $(< \sqrt{4\pi})$ 
& Neutrino Yukawa couplings \\
\hline


$\mathrm{BR}(\mu \to e\gamma)$ 
& $< 4.2 \times 10^{-13}$ 
& MEG experimental bound \\
\hline

$Y_{B}$ 
& $(8.7 \pm 0.1)\times10^{-11}$ 
& Observed baryon asymmetry \\
\hline

\end{tabular}
\end{table}

\subsection{High-scale hierarchical regime}

We first consider the standard thermal leptogenesis region with hierarchical right-handed neutrinos,
\begin{align}
10^9~{\rm GeV}\lesssim M_1 \lesssim 10^{13}~{\rm GeV}.
\end{align}
The numerical scan shows that successful baryogenesis strongly prefers larger values of $M_1$, consistent with the modified Davidson--Ibarra expectation in the scotogenic framework.

\begin{figure*}[!t]
\centering
\includegraphics[width=0.48\textwidth]{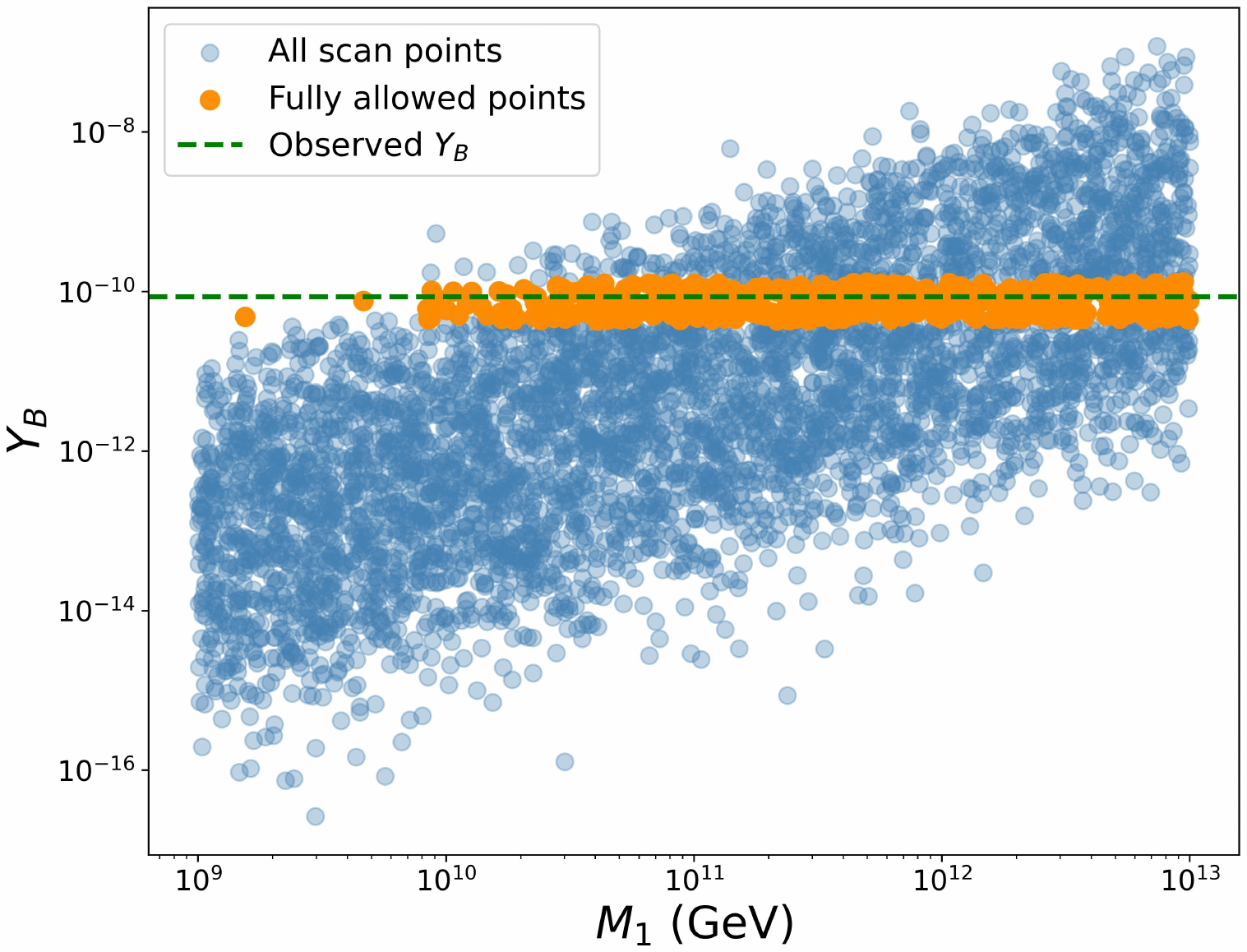}
\hfill
\includegraphics[width=0.48\textwidth]{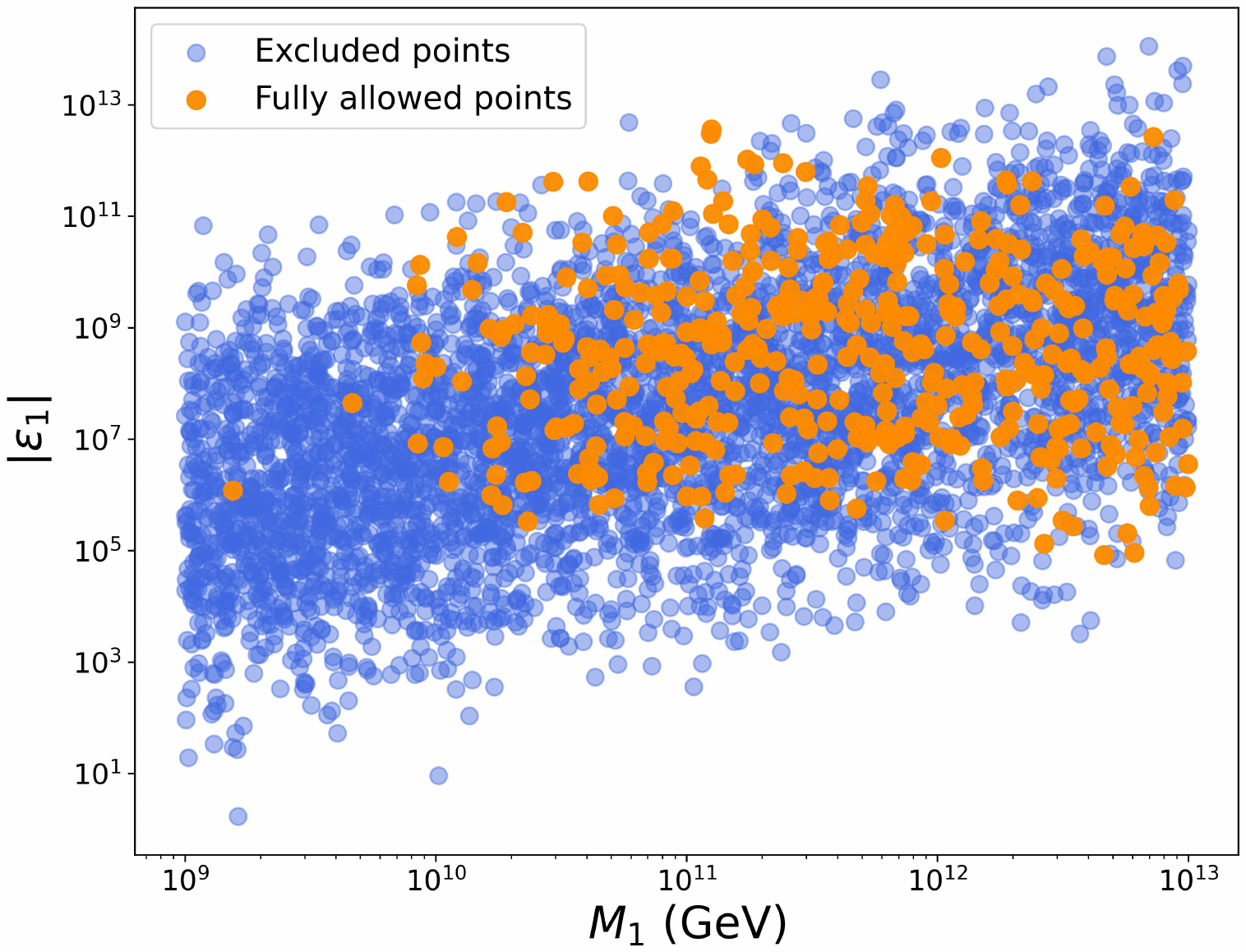}

\vspace{0.3cm}

\includegraphics[width=0.48\textwidth]{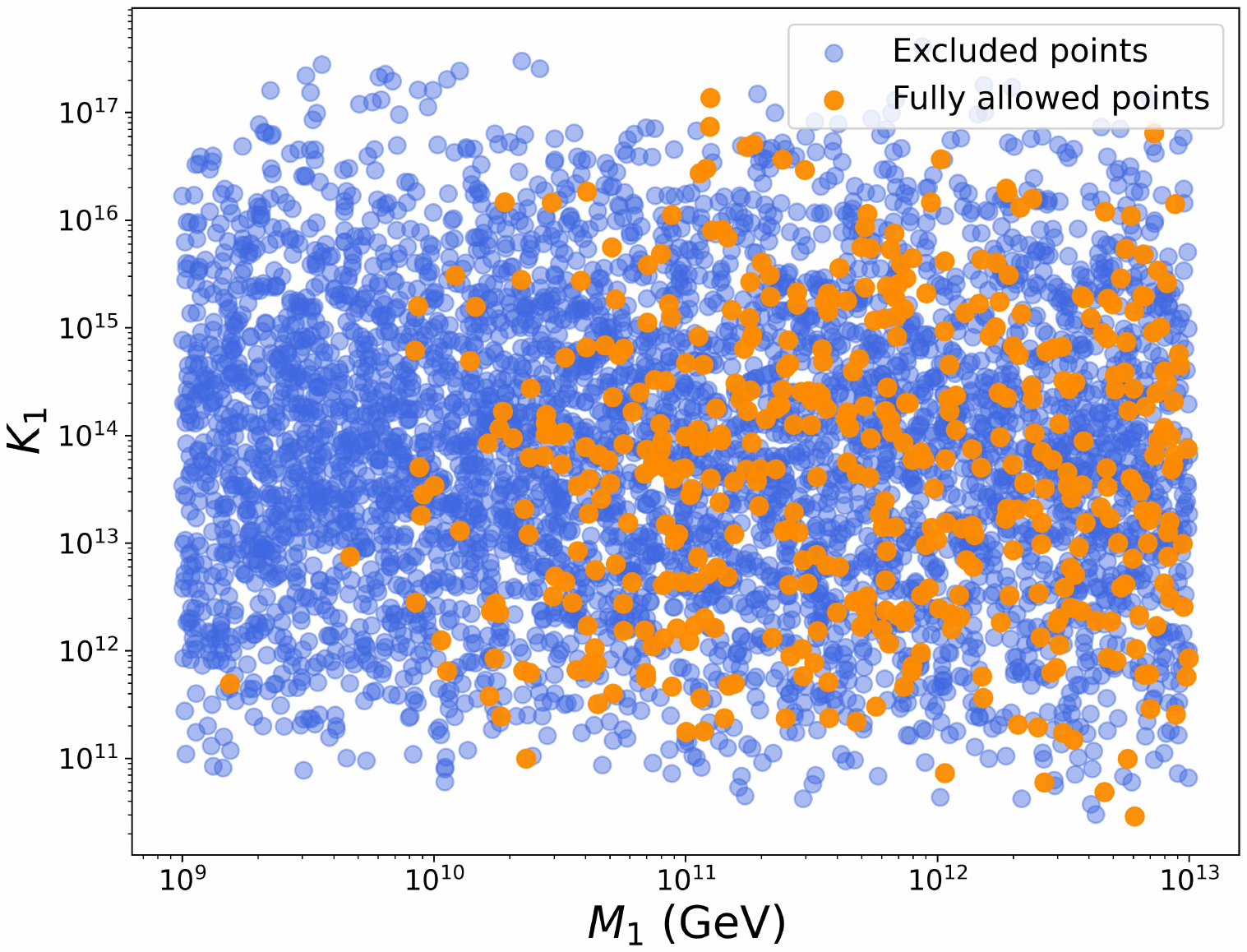}
\hfill
\includegraphics[width=0.48\textwidth]{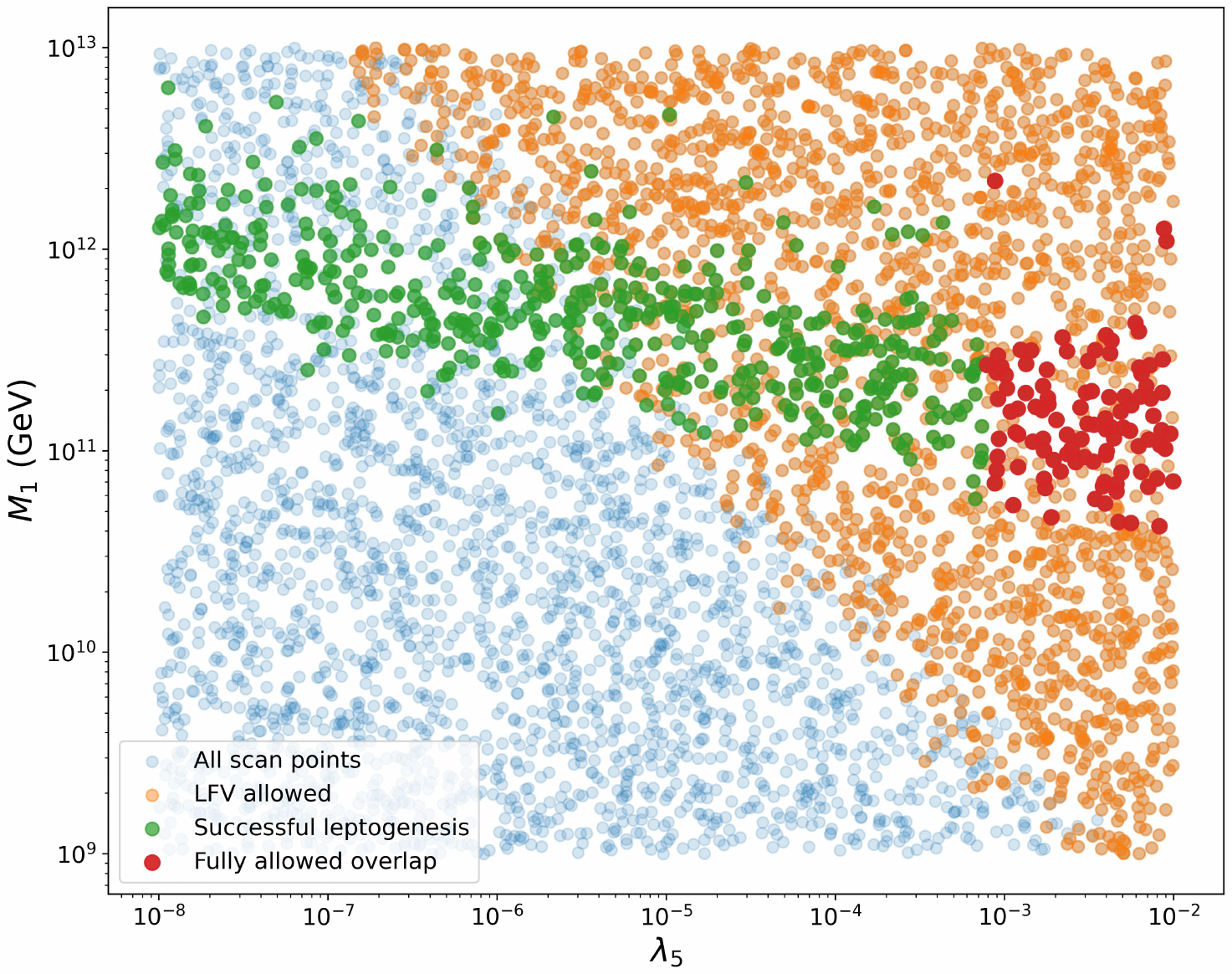}
\caption{
\textbf{High-scale hierarchical leptogenesis in the minimal scotogenic model.}
(\textbf{a}) Correlation between the baryon asymmetry $Y_B$ and the lightest RHN mass $M_1$. 
Successful leptogenesis requires sufficiently large RHN masses, typically in the high-scale regime 
$M_1 \gtrsim 10^{9}$--$10^{10}\,\mathrm{GeV}$, where the CP asymmetry becomes large enough to overcome washout suppression. 
The horizontal dashed line denotes the observed baryon asymmetry.
(\textbf{b}) CP asymmetry parameter $|\epsilon_1|$ as a function of $M_1$, showing the expected enhancement 
for heavier RHN masses consistent with the Davidson--Ibarra-type behavior.
(\textbf{c}) Washout parameter $K_1$ versus $M_1$. The viable region is concentrated in the intermediate-to-strong 
washout regime, indicating that successful baryogenesis is achieved without requiring fine-tuned weak washout solutions.
(\textbf{d}) Parameter-space distribution in the $(\lambda_5,\,M_1)$ plane. The fully allowed region is localized 
where neutrino mass generation, LFV constraints, and leptogenesis simultaneously remain consistent. 
The concentration of successful points confirms that high-scale leptogenesis is naturally realized in the strong washout regime.
}
\label{fig:highscale_main}
\end{figure*}

As shown in Fig.~\ref{fig:highscale_main}, the generated baryon asymmetry increases with $M_1$, and the observed value is typically reproduced for
\begin{align}
M_1 \gtrsim 10^{10}\text{--}10^{11}\ {\rm GeV},
\end{align}
depending on the Yukawa texture and the value of $\lambda_5$. This directly reflects the enhancement of the CP asymmetry,
\begin{align}
|\epsilon_1| \propto \frac{M_1}{\lambda_5},
\end{align}
which is clearly visible from the correlation between $|\epsilon_1|$ and $M_1$.

At the same time, the decay parameter $K_1$ remains large across the allowed region, confirming that the high-scale solution dominantly lies in the strong washout regime. This agrees with the analytical estimate
\begin{align}
K_1 \propto \frac{\widetilde m_1}{\lambda_5},
\end{align}
which explains why excessively small values of $\lambda_5$ become disfavored due to strong washout despite enhancing $\epsilon_1$.

The $(\lambda_5,M_1)$ plane reveals that viable points survive only in an intermediate window,
\begin{align}
10^{-3}\lesssim \lambda_5 \lesssim 10^{-2},
\end{align}
with
\begin{align}
M_1\sim 10^{11}\text{--}10^{12}\ {\rm GeV},
\end{align}
showing that maximizing CP asymmetry alone is insufficient; an optimal balance between asymmetry production and washout is required.

\begin{figure}[t]
\centering
\includegraphics[width=0.75\textwidth]{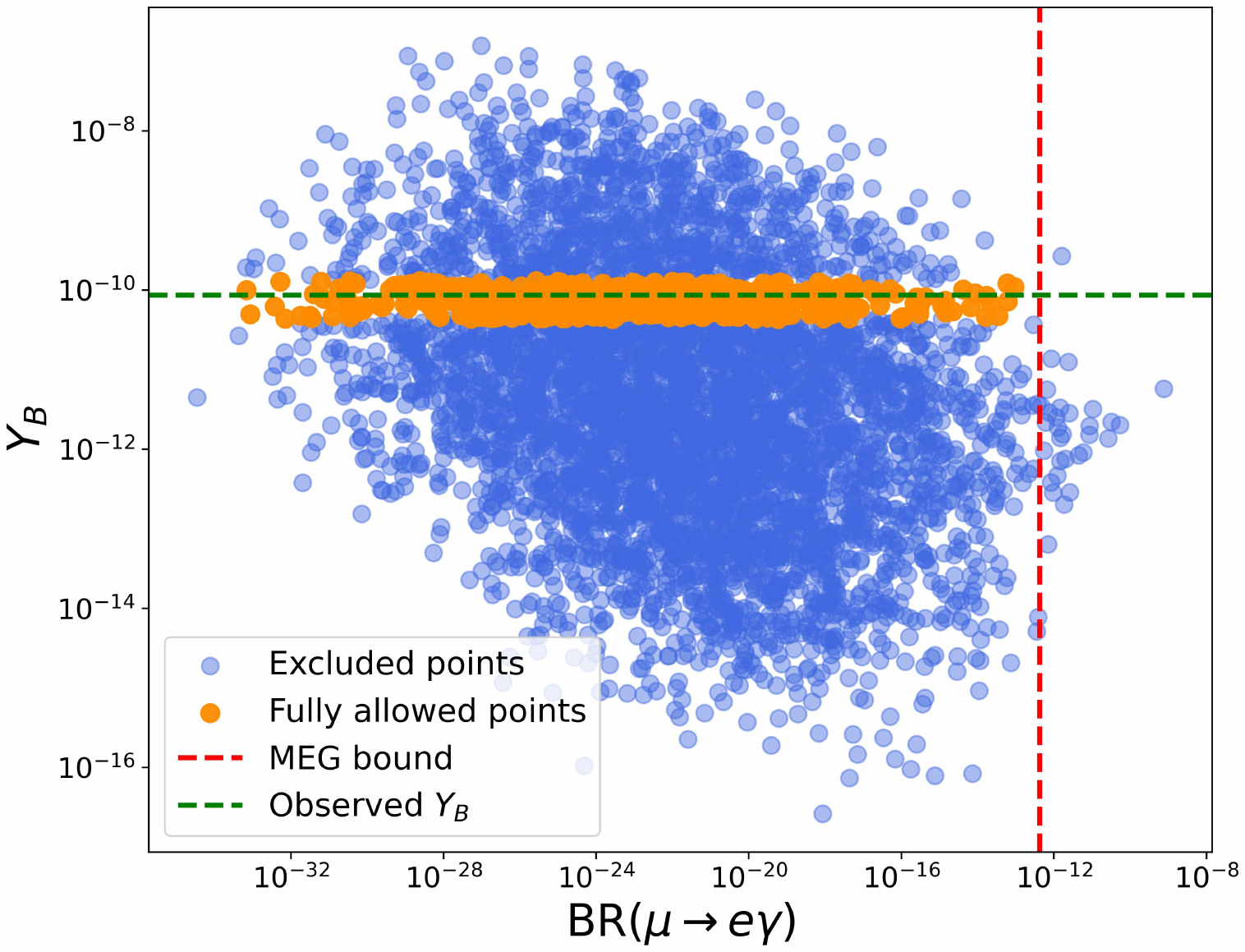}
\caption{
Interplay between lepton flavor violation and leptogenesis in the high-scale regime of the minimal scotogenic model. 
The horizontal green dashed line denotes the observed baryon asymmetry of the Universe,
$Y_B = (8.66 \pm 0.15)\times10^{-11},$
while the vertical red dashed line corresponds to the current experimental upper bound from the MEG collaboration,
$\mathrm{BR}(\mu \rightarrow e\gamma) < 4.2 \times 10^{-13}\text{at 90\% C.L.} $
Blue points represent excluded parameter-space points that fail either the LFV bound or the baryon asymmetry requirement, whereas orange points satisfy both successful leptogenesis and LFV constraints simultaneously. 
The narrow overlap region demonstrates that the same Yukawa structures controlling radiative neutrino mass generation also govern the tension between baryogenesis and charged lepton flavor violation, leading to a highly predictive allowed parameter space.
}
\label{fig:highscale_lfv}
\end{figure}

A strong correlation also emerges between LFV and leptogenesis. Figure~\ref{fig:highscale_lfv} shows that the MEG bound removes a significant part of the parameter space by constraining the same Yukawa combinations entering leptogenesis. The overlap region therefore becomes highly predictive and considerably sharpens the viable benchmark structure.

\subsection{Low-scale quasi-degenerate regime}

We next study the low-scale resonant leptogenesis scenario where the right-handed neutrinos are quasi-degenerate and the self-energy contribution enhances the CP asymmetry, allowing successful baryogenesis at much smaller masses,
\begin{align}
10^5~{\rm GeV}\lesssim M_1 \lesssim 10^7~{\rm GeV}.
\end{align}

\begin{figure*}[!t]
\centering

\includegraphics[width=0.48\textwidth]{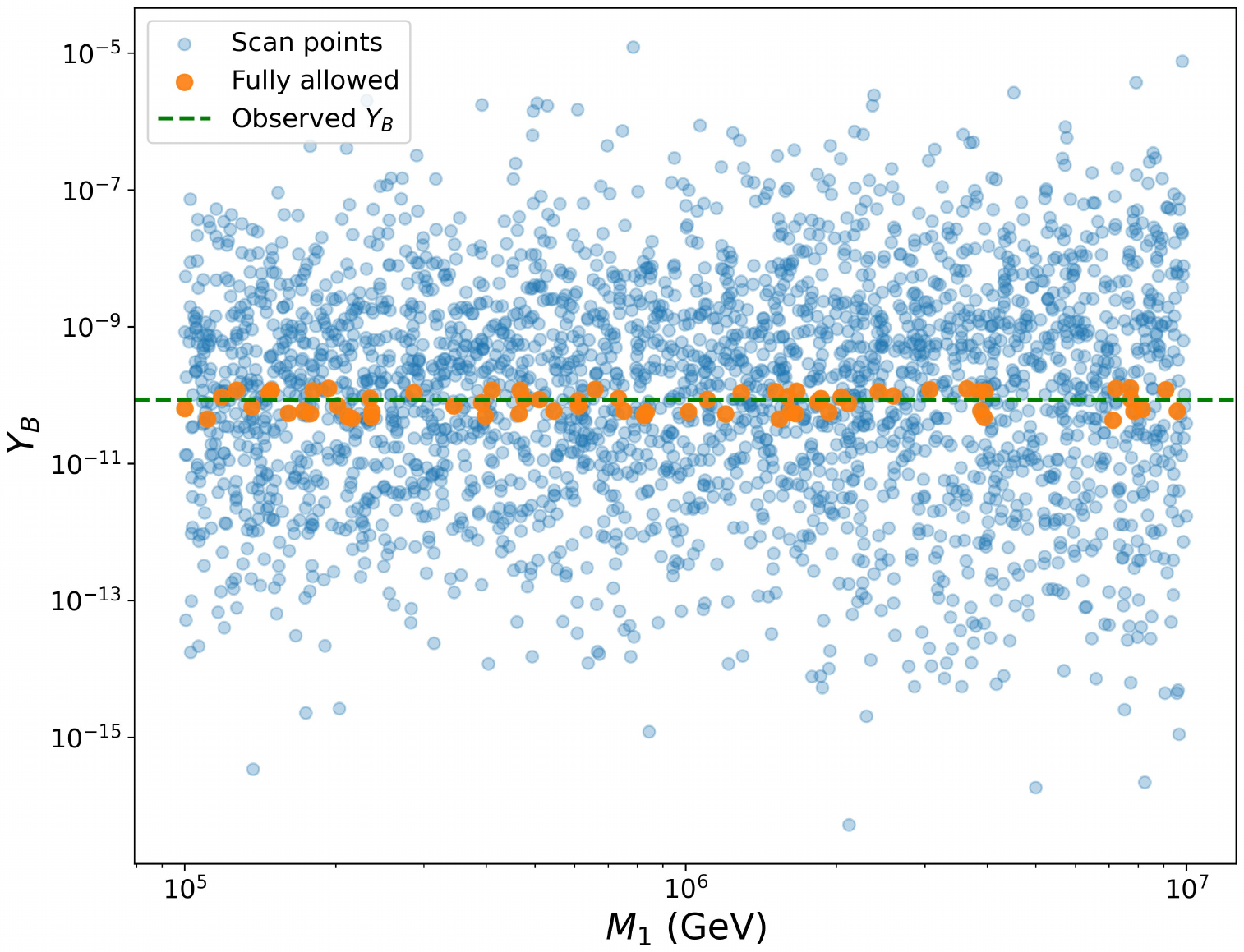}
\hfill
\includegraphics[width=0.48\textwidth]{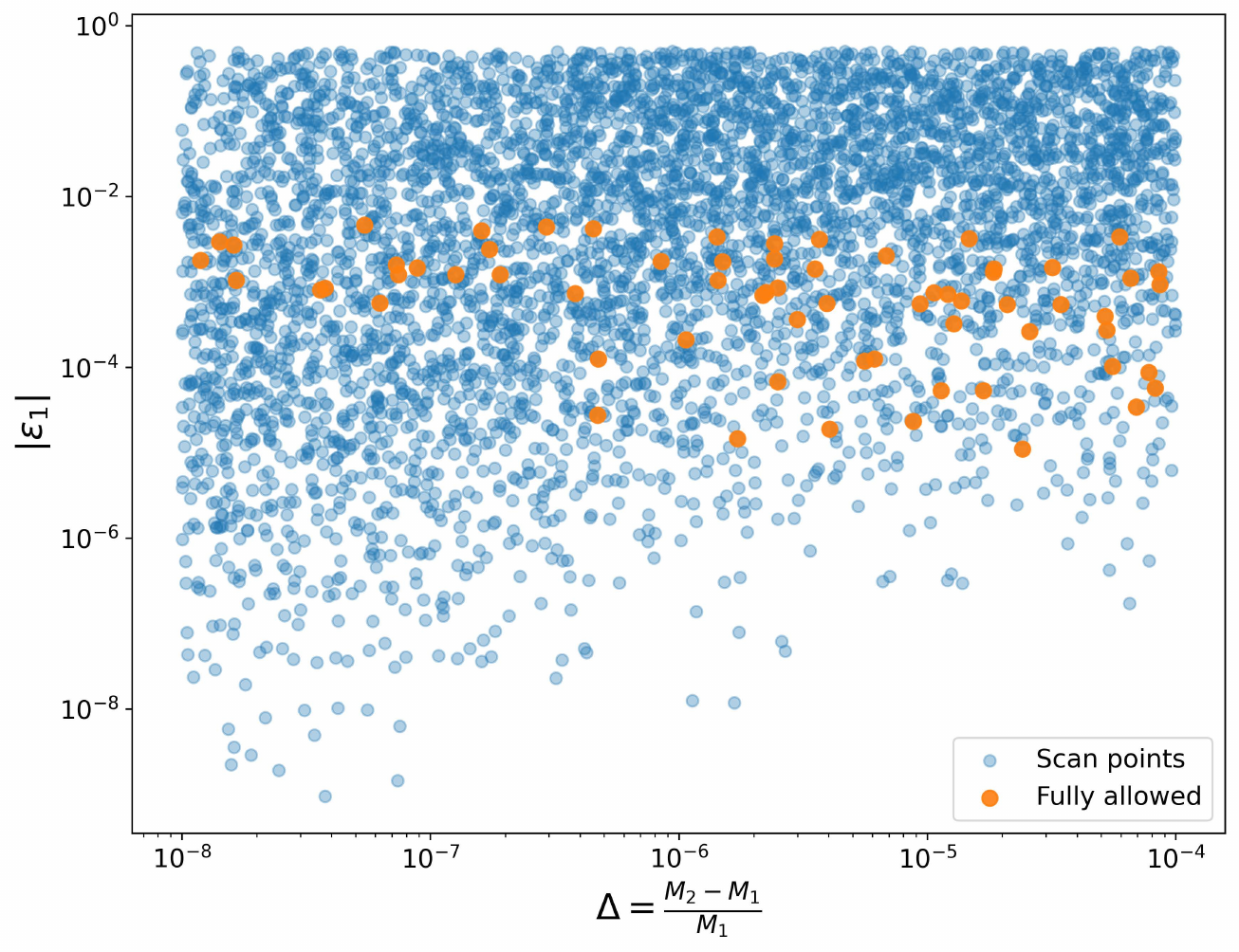}

\vspace{0.3cm}

\includegraphics[width=0.48\textwidth]{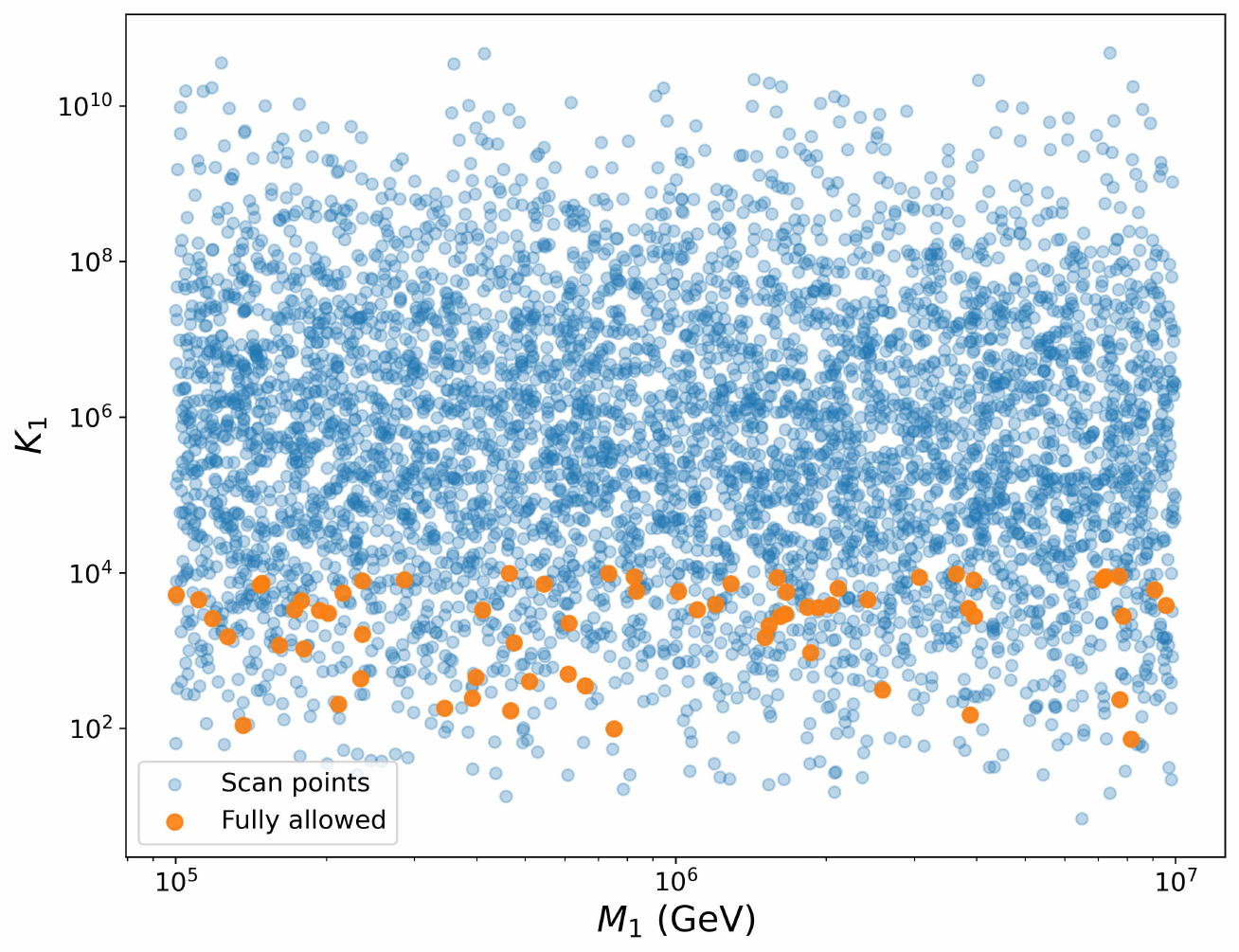}
\hfill
\includegraphics[width=0.48\textwidth]{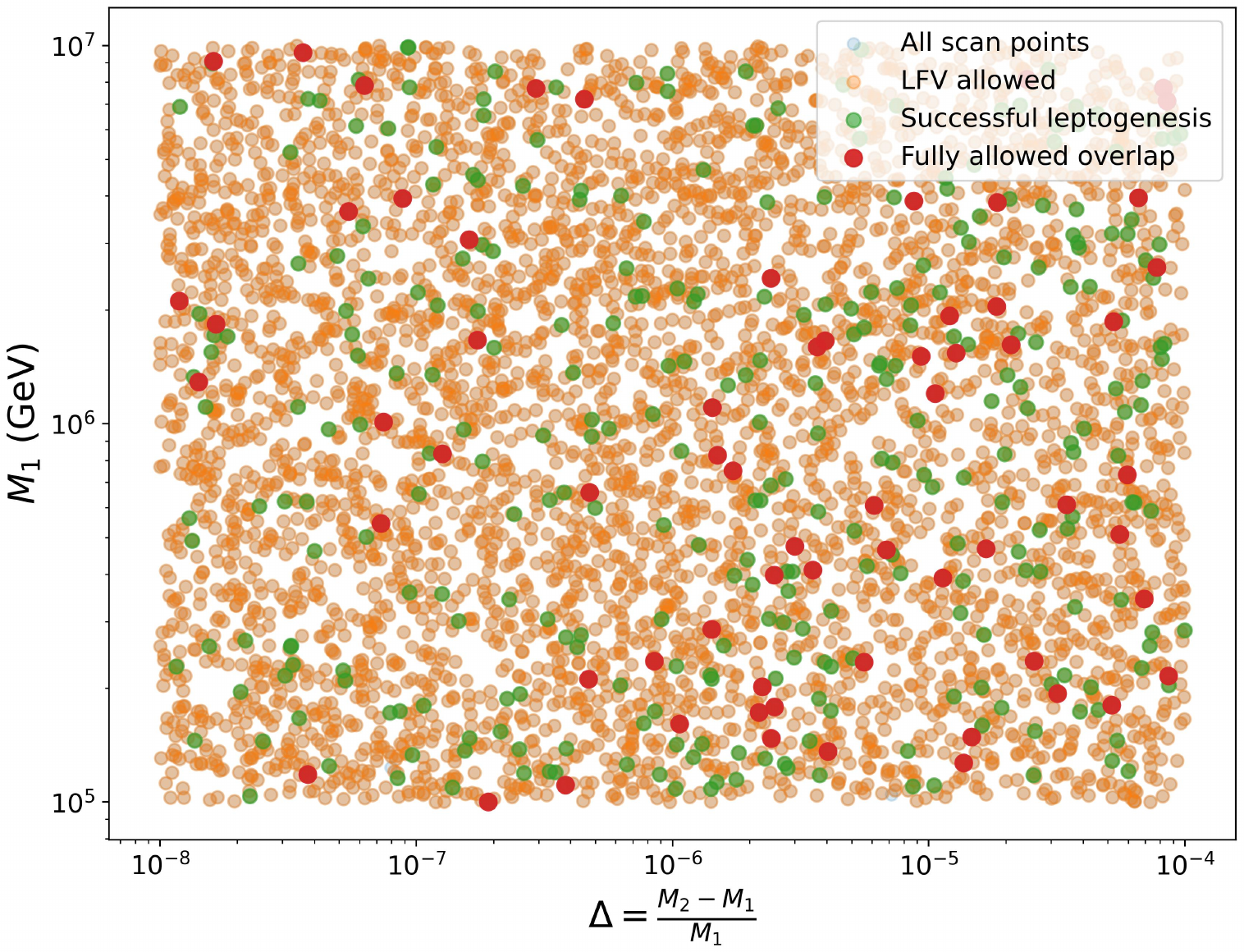}
\caption{
\textbf{Low-scale resonant leptogenesis with quasi-degenerate right-handed neutrinos in the minimal scotogenic model.}
(\textbf{a}) Baryon asymmetry $Y_B$ as a function of the lightest RHN mass $M_1$. 
The horizontal dashed line denotes the observed baryon asymmetry of the Universe. 
Successful leptogenesis is achieved for comparatively low RHN masses once resonant enhancement is present, allowing viable baryogenesis far below the conventional hierarchical leptogenesis scale.
(\textbf{b}) CP asymmetry parameter $|\epsilon_1|$ as a function of the relative mass splitting
$\Delta = \frac{M_2-M_1}{M_1}.$
The enhancement of $|\epsilon_1|$ for small $\Delta$ clearly demonstrates the resonant origin of the baryon asymmetry, with the viable region concentrated near the quasi-degenerate limit.
(\textbf{c}) Washout parameter $K_1$ versus $M_1$. 
The fully allowed points are localized in the intermediate-to-strong washout regime,
typically
$10^{2} \lesssim K_1 \lesssim 10^{4},$
showing that successful leptogenesis requires a balance between sufficient CP asymmetry and controlled inverse-decay suppression.
(\textbf{d}) Distribution of scan points in the $(\Delta,\,M_1)$ plane. 
The fully allowed region is strongly localized in a narrow resonant strip where the RHN mass splitting is sufficiently small to compensate the strong washout while remaining consistent with LFV constraints and neutrino mass generation. 
This demonstrates the strong predictivity of low-scale resonant leptogenesis in the minimal scotogenic framework.
}
\label{fig:lowscale_main}
\end{figure*}

The crucial parameter in this case is the mass degeneracy
\begin{align}
\Delta=\frac{M_2-M_1}{M_1}.
\end{align}
Figure~\ref{fig:lowscale_main} shows that successful leptogenesis requires
\begin{align}
10^{-6} \leq \Delta \leq 10^{-2}
\end{align}
fully allowed points cluster around
$10^{-4} \lesssim \Delta \lesssim 10^{-2}$ where resonant enhancement becomes efficient. This is fully consistent with the analytical expectation that the CP asymmetry grows significantly as the RHN masses approach degeneracy.

Unlike the high-scale case, the allowed values of $K_1$ are substantially smaller and populate the moderate washout regime,
\begin{align}
10^2\lesssim K_1 \lesssim 10^4,
\end{align}
which improves the efficiency factor and allows the observed baryon asymmetry to be reproduced even for relatively small $M_1$. This confirms that resonant enhancement relaxes the conventional thermal lower bound on the leptogenesis scale.

The preferred range of the scalar coupling is found to be
\begin{align}
10^{-4}\lesssim \lambda_5 \lesssim 10^{-2},
\end{align}
where neutrino mass generation and leptogenesis can be simultaneously maintained. Very small $\lambda_5$ again leads to excessive washout, while very large values suppress the loop-induced neutrino masses and reduce the efficiency of asymmetry generation.

The final baryon asymmetry distribution demonstrates that once the resonance condition is satisfied, the observed value of $Y_B$ can be reproduced throughout the low-scale window, establishing a fully viable alternative to the conventional high-scale scenario.

\begin{figure}[t]
\centering
\includegraphics[width=0.48\textwidth]{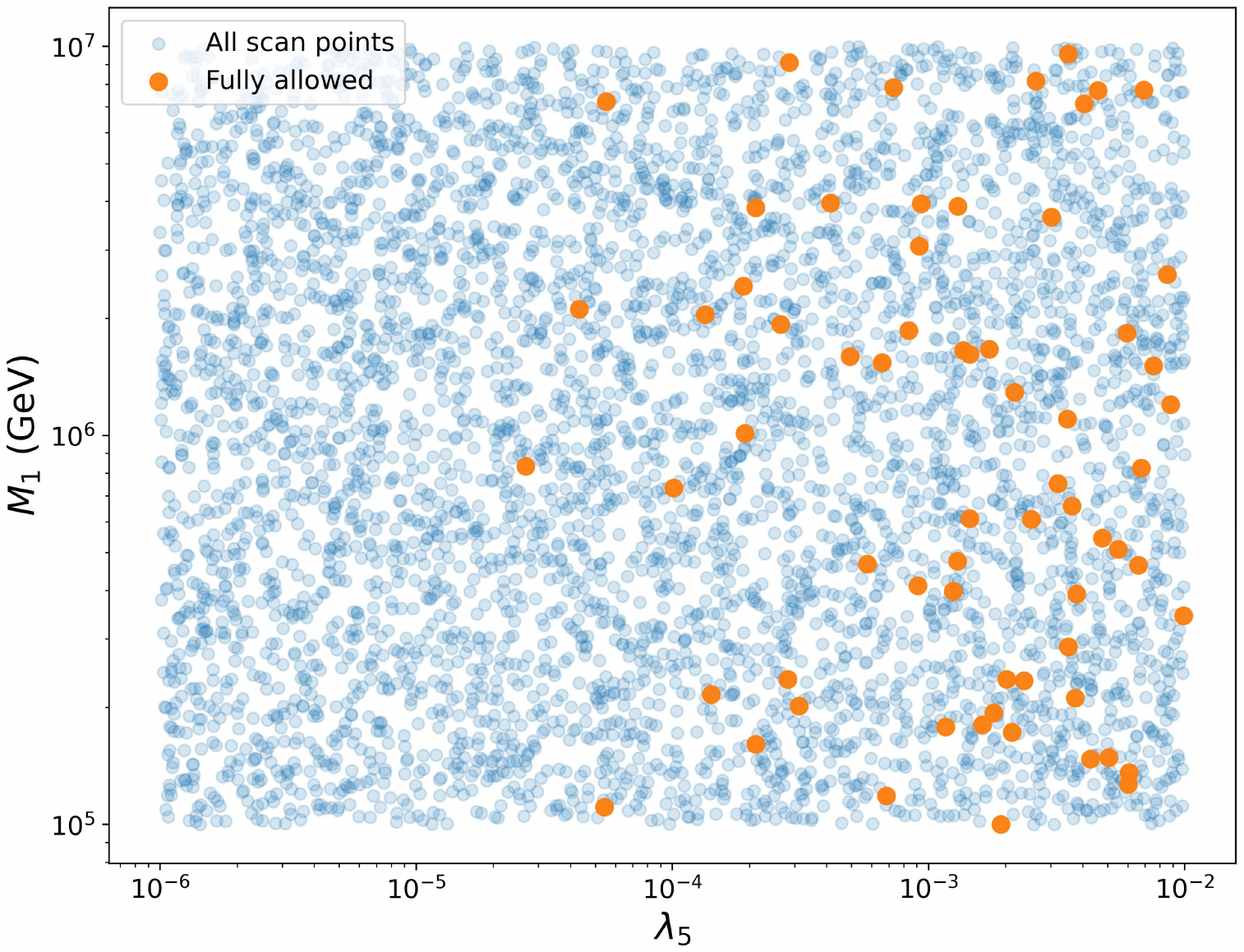}
\hfill
\includegraphics[width=0.48\textwidth]{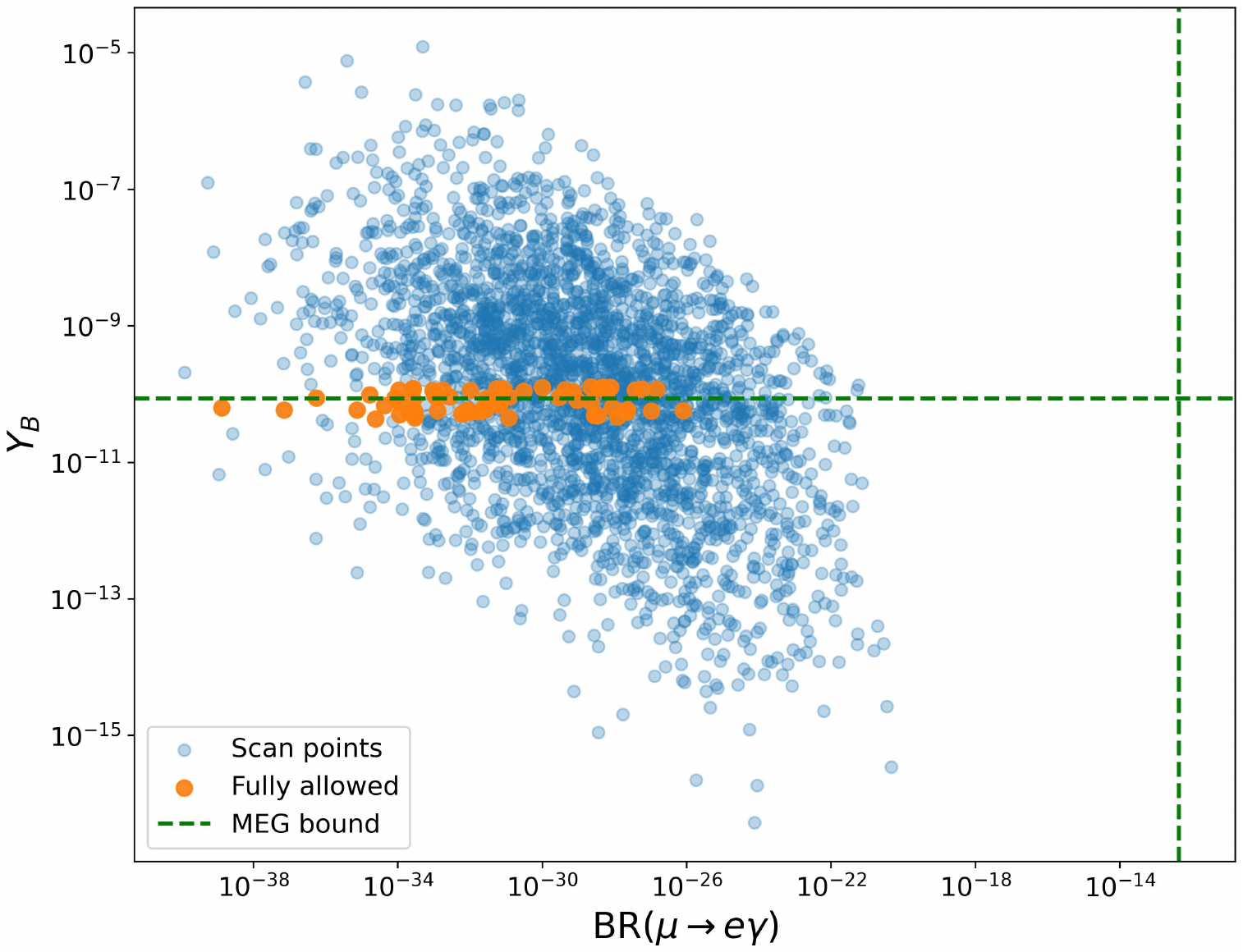}
\caption{
\textbf{Low-scale resonant parameter space after imposing leptogenesis and LFV constraints.}
(\textbf{a}) Distribution of scan points in the $(\lambda_5,\,M_1)$ plane. 
The fully allowed points are concentrated in a restricted region with moderate values of the scalar coupling $\lambda_5$ and sufficiently low RHN masses, reflecting the nontrivial interplay between radiative neutrino mass generation, perturbativity, and resonant leptogenesis. 
Very small values of $\lambda_5$ enhance the CP asymmetry but simultaneously increase the washout parameter $K_1$, while excessively large values suppress the baryon asymmetry, leading to a preferred intermediate window.
(\textbf{b}) Correlation between the baryon asymmetry $Y_B$ and the branching ratio
$
\mathrm{BR}(\mu \rightarrow e\gamma).
$
The horizontal dashed line denotes the observed baryon asymmetry of the Universe, while the vertical dashed line corresponds to the current MEG upper bound on charged lepton flavor violation. 
The surviving orange points indicate the narrow LFV-safe resonant strip where successful leptogenesis and LFV consistency coexist simultaneously. This overlap region provides one of the sharpest experimentally testable predictions of the minimal scotogenic model.
}
\label{fig:lowscale_overlap}
\end{figure}

\begin{table}[t]
\centering
\caption{Representative benchmark points for low-scale and high-scale leptogenesis satisfying all phenomenological constraints.}
\label{tab:benchmark_points}
\begin{tabular}{|c|c|c|}
\hline
Parameter & Low-scale Benchmark & High-scale Benchmark \\
\hline
$M_{1}$ (GeV) & $5.2 \times 10^{5}$ & $3.4 \times 10^{10}$ \\
\hline
$M_{2}$ (GeV) & $5.20003 \times 10^{5}$ & $3.40002 \times 10^{10}$ \\
\hline
$\Delta = \dfrac{M_{2}-M_{1}}{M_{1}}$ & $5.8 \times 10^{-6}$ & hierarchical \\
\hline
$\lambda_{5}$ & $2.3 \times 10^{-4}$ & $7.6 \times 10^{-3}$ \\
\hline
$\theta_{1}$ & $1.72$ & $4.91$ \\
\hline
$\theta_{2}$ & $3.84$ & $2.17$ \\
\hline
$|\varepsilon_{1}|$ & $3.8 \times 10^{-5}$ & $1.6 \times 10^{-6}$ \\
\hline
$K_{1}$ & $ 10^{2}\text{--}10^{4}$ & $8.7 \times 10^{3}$ \\
\hline
$\mathrm{BR}(\mu \to e\gamma)$ & $1.8 \times 10^{-13}$ & $4.5 \times 10^{-18}$ \\
\hline
$Y_{B}$ & $8.6 \times 10^{-11}$ & $8.9 \times 10^{-11}$ \\
\hline
\end{tabular}
\end{table}


The benchmark points clearly demonstrate the complementarity between the two leptogenesis regimes. In the low-scale case, the branching ratio for
\[
\mu \rightarrow e\gamma
\]
lies close to the current MEG sensitivity, making it phenomenologically attractive and potentially testable in near-future experiments. The corresponding strong washout parameter
\[
K_{1} \gg 1
\]
is compensated by a sizable resonantly enhanced CP asymmetry, leading to the correct baryon asymmetry.

In contrast, the high-scale benchmark exhibits strongly suppressed LFV rates due to the larger heavy neutrino mass scale, while still maintaining successful leptogenesis through conventional thermal production. Although experimentally less accessible in LFV searches, this regime provides an important consistency check of the model across a broad mass range.

These benchmark points therefore confirm that the model accommodates both experimentally testable low-scale leptogenesis and theoretically robust high-scale leptogenesis within the same unified framework.

\subsection{Comparison with analytical expectations}

The numerical results strongly support the analytical discussion presented earlier.

First, the scaling
\begin{align}
\epsilon_1 \sim \frac{M_1}{\lambda_5}
\end{align}
is directly confirmed in the high-scale scan. Second, the approximate lower bound
\begin{align}
K_1 \gtrsim \mathcal{O}(10^3)
\end{align}
naturally explains why the hierarchical regime is dominated by strong washout. Third, the existence of an optimal intermediate range of $\lambda_5$ verifies that maximizing CP asymmetry alone does not maximize the final baryon asymmetry; washout effects must be simultaneously controlled.

In the low-scale regime, quasi-degeneracy replaces the Davidson--Ibarra bound through resonant enhancement and allows viable leptogenesis down to
\begin{align}
M_1\sim 10^5~{\rm GeV},
\end{align}
providing a non-trivial consistency check of the analytical framework.

Overall, the minimal scotogenic model exhibits two robust and complementary viable regions: a high-scale hierarchical solution characterized by strong washout and a low-scale quasi-degenerate solution driven by resonant enhancement. The simultaneous inclusion of leptogenesis, LFV, and neutrino mass constraints significantly sharpens the parameter space and yields well-defined benchmark regions for precision phenomenological studies.

\section{Conclusions}

In this work, we have performed a systematic study of the interplay between lepton
flavor violation (LFV) and leptogenesis in the minimal scotogenic model, focusing on
the simultaneous realization of radiative neutrino mass generation, successful baryogenesis,
and compatibility with present flavor constraints. To our knowledge, this is among the first systematic studies
directly comparing high-scale hierarchical leptogenesis and
low-scale resonant leptogenesis under present LFV constraints
within the same full Casas–Ibarra reconstructed scotogenic framework.

The scotogenic framework is particularly predictive because the same Yukawa couplings
that generate neutrino masses at one loop also control charged lepton flavor violating
processes such as $\mu \to e\gamma$ and the CP asymmetry responsible for leptogenesis.
This creates a strong and highly nontrivial correlation between neutrino physics, flavor
observables, and early-Universe baryogenesis. The benchmark points show that both high-scale and low-scale
leptogenesis remain compatible with present cosmological limits
on $\sum m_\nu$ and predict small but nonzero values of
$m_{ee}$, characteristic of normal ordering in the scotogenic framework.

Using the full Casas--Ibarra reconstruction of the Yukawa sector, we investigated both
the high-scale hierarchical leptogenesis regime and the low-scale resonant leptogenesis
regime within a unified framework. We included the latest neutrino oscillation data,
the MEG bound on
\[
\mathrm{BR}(\mu \to e\gamma) < 4.2 \times 10^{-13},
\]
and the observed baryon asymmetry of the Universe as the main phenomenological
constraints. MEG II can directly test the surviving strip that we have presented.

For the high-scale hierarchical regime, we found that successful leptogenesis naturally
survives for
\[
M_1 \gtrsim 10^{10}\text{--}10^{11}\,\mathrm{GeV},
\]
with an intermediate range of the scalar coupling $\lambda_5$. In this region, the CP
asymmetry follows the modified Davidson--Ibarra behavior, while the system typically
lies in the strong washout regime with large values of the decay parameter $K_1$.
Importantly, LFV constraints do not exclude this scenario because flavor alignment and
Casas--Ibarra phase cancellations allow an effective decoupling between baryogenesis
and low-energy flavor violation.

The low-scale quasi-degenerate regime exhibits a qualitatively different structure.
Here, resonant enhancement generated by the quasi-degeneracy condition
\[
M_2 \simeq M_1
\]
allows successful leptogenesis at much lower scales,
\[
M_1 \sim 10^{5}\text{--}10^{7}\,\mathrm{GeV},
\]
far below the conventional thermal leptogenesis bound. However, this regime is strongly
constrained because the same Yukawa enhancement required for resonant CP asymmetry
also tends to increase both LFV amplitudes and washout effects. As a result, most of
the parameter space is excluded by the MEG bound and excessive inverse-decay washout.

Nevertheless, contrary to the common expectation of complete exclusion, we identified
a narrow but nonvanishing resonant window in which successful baryogenesis, controlled
washout, and LFV safety coexist simultaneously. This surviving region is characterized
by quasi-degenerate heavy fermions, moderate values of $\lambda_5$, nonzero complex
Casas--Ibarra phases, and suppressed flavor-violating amplitudes through phase
alignment. Fully viable benchmark points were obtained with perturbative Yukawa
couplings and resonantly enhanced CP asymmetry.

Our analytical estimates for the CP asymmetry, washout parameter, and final baryon
asymmetry are in excellent agreement with the numerical scan and provide a transparent
physical interpretation of the allowed parameter space. In particular, we find that
maximizing the CP asymmetry alone does not maximize the final baryon asymmetry;
instead, successful leptogenesis requires an optimal balance between asymmetry
generation and washout suppression.

The central result of this work may therefore be summarized as follows:

\begin{center}
\emph{
High-scale leptogenesis survives naturally, whereas low-scale resonant leptogenesis
survives only in a highly restricted LFV-safe resonant strip.
}
\end{center}

This conclusion provides a strong phenomenological target for future charged lepton
flavor violation searches, especially for MEG II, Mu3e, Mu2e, and COMET, whose
improved sensitivities can probe a significant fraction of the remaining low-scale
parameter space.

Although the scotogenic model also offers a viable dark matter candidate through the
lightest $Z_2$-odd particle, a full dark matter relic density and direct-detection
analysis was beyond the scope of the present work. Such an extension would provide
an even stronger test of the surviving resonant window and constitutes an important
direction for future investigation.

Overall, the minimal scotogenic model continues to provide one of the most economical
and predictive frameworks linking neutrino masses, dark matter, lepton flavor violation,
and baryogenesis. The simultaneous inclusion of all these sectors significantly sharpens
the viable parameter space and makes the model highly testable in upcoming experiments. The surviving low-scale resonant strip predicts
$\mathrm{BR}(\mu \to e\gamma)$ close to the present
MEG sensitivity and can be directly tested by the upcoming
MEG II experiment, while complementary searches such as
Mu3e, Mu2e, and COMET may further probe the remaining
parameter space.

\appendix

\section{Loop functions and benchmark Yukawa structures}

In this appendix, we collect the loop functions relevant for neutrino mass generation,
charged lepton flavor violation, and leptogenesis in the minimal scotogenic model.
We also present representative benchmark Yukawa textures corresponding to the viable
regions identified in the numerical analysis.


\subsection{One-loop neutrino mass function}

In the scotogenic model, the light neutrino mass matrix is generated radiatively
through the exchange of the inert scalar doublet and the singlet fermions \(N_i\).
The neutrino mass matrix is given by
\begin{equation}
(M_\nu)_{\alpha\beta}
=
\sum_{i=1}^{3}
\frac{Y_{\alpha i}Y_{\beta i}M_i}{32\pi^2}
\left[
\frac{m_R^2}{m_R^2-M_i^2}
\ln\left(\frac{m_R^2}{M_i^2}\right)
-
\frac{m_I^2}{m_I^2-M_i^2}
\ln\left(\frac{m_I^2}{M_i^2}\right)
\right],
\label{eq:appendix_neutrino_mass}
\end{equation}
where \(m_R\) and \(m_I\) denote the CP-even and CP-odd neutral inert scalar masses,
respectively.

For small scalar splitting,
\begin{equation}
m_R^2 - m_I^2
=
\lambda_5 v^2,
\end{equation}
the above expression can be rewritten as
\begin{equation}
(M_\nu)_{\alpha\beta}
=
\sum_{i=1}^{3}
Y_{\alpha i}Y_{\beta i}\Lambda_i,
\end{equation}
with
\begin{equation}
\Lambda_i
=
\frac{\lambda_5 v^2}{32\pi^2}
\frac{M_i}{m_\eta^2-M_i^2}
\left[
1
-
\frac{M_i^2}{m_\eta^2-M_i^2}
\ln\left(\frac{m_\eta^2}{M_i^2}\right)
\right].
\label{eq:lambda_i_appendix}
\end{equation}

This form is particularly useful for the Casas--Ibarra reconstruction.


\subsection{Loop function for \(\mu \to e\gamma\)}

The branching ratio for the LFV decay
\begin{equation}
\mu \to e\gamma
\end{equation}
depends on the loop function
\begin{equation}
F_2(x)
=
\frac{
1
-
6x
+
3x^2
+
2x^3
-
6x^2\ln x
}{
6(1-x)^4
},
\label{eq:F2_appendix}
\end{equation}
where
\begin{equation}
x_i
=
\frac{M_i^2}{m_{\eta^\pm}^2}.
\end{equation}

The corresponding branching ratio is
\begin{equation}
\mathrm{BR}(\mu \to e\gamma)
=
\frac{
3\alpha_{\rm em}
}{
64\pi
G_F^2
m_{\eta^\pm}^4
}
\left|
\sum_i
Y_{\mu i}
Y_{e i}^*
F_2(x_i)
\right|^2.
\label{eq:BR_appendix}
\end{equation}

This expression shows explicitly that the same Yukawa couplings responsible for
radiative neutrino masses also control LFV observables.


\subsection{CP asymmetry loop function}

The CP asymmetry generated in the decay of the lightest singlet fermion \(N_1\)
receives both vertex and self-energy contributions. It is given by
\begin{equation}
\epsilon_1
=
\frac{
1
}{
8\pi
(Y^\dagger Y)_{11}
}
\sum_{j\neq1}
\mathrm{Im}
\left[
\left(
Y^\dagger Y
\right)^2_{1j}
\right]
f\!\left(
\frac{M_j^2}{M_1^2}
\right),
\label{eq:eps_appendix}
\end{equation}
where the loop function can be written as
\begin{equation}
f(r)
=
\sqrt{r}
\left[
\frac{1}{1-r}
+
1
-
(1+r)\ln\left(\frac{1+r}{r}\right)
\right].
\label{eq:f_lepto_appendix}
\end{equation}

In the quasi-degenerate limit
\begin{equation}
M_2
\simeq
M_1,
\end{equation}
the self-energy contribution becomes resonantly enhanced and dominates the baryon
asymmetry generation.



\end{document}